\documentclass{v16nufact-oy}

\def\be{\begin{equation}}
\def\ee{\end{equation}}
\def\bea{\begin{eqnarray}}
\def\eea{\end{eqnarray}}

\newcommand{\lesssim}{\raisebox{0.3mm}{\em $\, <$} \hspace{-3.2mm}
\raisebox{-1.4mm}{\em $\sim \,$}}
\newcommand{\Photo}{}

\begin{document}
\vspace*{4cm}
\title{Will atmospheric neutrino experiment at Hyper-Kamiokande see
non-standard interaction effects?}

\author{Osamu Yasuda}

\address{Department of Physics, Tokyo Metropolitan University,\\
 Hachioji, Tokyo 192-0397, Japan}

\maketitle\abstracts{ In this talk we discuss the possibility to test
  the hypothesis, which has been proposed to explain the tension
  between the mass-squared differences of the solar
  neutrino and KamLAND experiments by the non-standard
  flavor-dependent interaction in neutrino propagation, with the atmospheric
  neutrino observations at the future Hyper-Kamiokande experiment.  }

\section{Introduction}
In the standard three flavor framework,
neutrino oscillations are described by
the mixing matrix
\begin{eqnarray}
U=\left(
\begin{array}{ccc}
c_{12}c_{13} & s_{12}c_{13} &  s_{13}e^{-i\delta}\nonumber\\
-s_{12}c_{23}-c_{12}s_{23}s_{13}e^{i\delta} & 
c_{12}c_{23}-s_{12}s_{23}s_{13}e^{i\delta} & s_{23}c_{13}\nonumber\\
s_{12}s_{23}-c_{12}c_{23}s_{13}e^{i\delta} & 
-c_{12}s_{23}-s_{12}c_{23}s_{13}e^{i\delta} & c_{23}c_{13}\nonumber\\
\end{array}\right),
\label{eqn:mns}
\end{eqnarray}
where the notations are as follows:
$c_{ij}\equiv\cos\theta_{ij}$,
$s_{ij}\equiv\sin\theta_{ij}$ and 
$\theta_{ij}$~$(j,k)=(1,2), (1,3), (2,3)$
are the three mixing angles and $\delta$ is
the CP phase.
Thanks to the successful results of the
solar, atmospheric, reactor and accelerator neutrino
experiments,
the three mixing angles and the two mass squared differences
have been measured.  The currently unknown quantities
are the mass hierarchy pattern (sign($\Delta m^2_{31}$)), 
the octant of $\theta_{23}$  (sign($\pi/4 - \theta_{23}$))
and $\delta$.
It is expected that these unknown quantities will be
measured by the neutrino experiments in the future,
particularly those with intense accelerator neutrino
beams\,\cite{Abe:2014oxa,Acciarri:2015uup}.
These future experiments are expected to
probe new physics beyond the
standard model with massive neutrinos,
from the deviation from the standard scheme.

In the standard three flavor framework of neutrinos,
the Dirac equation for the flavor eigenstate
$\Psi^T\equiv(\nu_e,\nu_\mu,\nu_\tau)$ of neutrino in matter
is given by
\begin{eqnarray}
i{d\Psi \over dt}=
\left[U \mbox{\rm diag}\left(E_1,E_2,E_3\right) U^{-1}
+{\cal A}
\right]\Psi\,.
\label{sch1}
\end{eqnarray}
Here the matter potential ${\cal A}$ is given by
\begin{eqnarray}
{\cal A}=A\left(
\begin{array}{ccc}
1 & 0 & 0\\
0 & 0 & 0\\
0 & 0 & 0
\end{array}
\right).
\label{matter-std}
\end{eqnarray}
$A\equiv \sqrt{2} G_F n_e$
stands for the standard matter effect which
comes from the
charged current interaction,
$n_e$ is the number density of the electron in the matter.

It was pointed out in Ref.\,\refcite{Gonzalez-Garcia:2013usa} that there is a
tension between the mass-squared difference deduced from the
solar neutrino observations and the one from the KamLAND experiment, and that the tension can be
resolved by introducing the flavor-dependent NSI
in neutrino propagation.
Such a hint for NSI gives us a strong motivation to study NSI in propagation in details.\footnote{
Some models predict large non-standard
interactions\,\cite{Farzan:2016now}, and
hence such large NSI effects are worth investigating also from the
view point of model building.}

The flavor-dependent nonstandard four-fermi interactions
which is discussed in this talk are given by
\begin{eqnarray}
{\cal L}_{\mbox{\rm\scriptsize eff}}^{\mbox{\rm\scriptsize NSI}} =
-2\sqrt{2}\, \epsilon_{\alpha\beta}^{fP} G_F
(\overline{\nu}_\alpha \gamma_\mu P_L \nu_\beta)\,
(\overline{f} \gamma^\mu P f),
\label{NSIop}
\end{eqnarray}
where only the interactions with $f = e, u, d$ are relevant to
the flavor transition of neutrino due to the matter effect,
$G_F$ denotes the Fermi coupling constant, $P$ stands for
a projection operator and is either
$P_L\equiv (1-\gamma_5)/2$ or $P_R\equiv (1+\gamma_5)/2$.
In the presence of these interactions (\ref{NSIop}),
the matter potential is modified to
\begin{eqnarray}
{\cal A} = A\left(
\begin{array}{ccc}
1+ \epsilon_{ee} & \epsilon_{e\mu} & \epsilon_{e\tau}\\
\epsilon_{e\mu}^\ast & \epsilon_{\mu\mu} & \epsilon_{\mu\tau}\\
\epsilon_{e\tau}^\ast & \epsilon_{\mu\tau}^\ast & \epsilon_{\tau\tau}
\end{array}
\right),
\label{matter-nsi}
\end{eqnarray}
where 
$\epsilon_{\alpha\beta}$ are defined as
$\epsilon_{\alpha\beta}
\equiv \sum_{f,P}(n_f/n_e) \epsilon_{\alpha\beta}^{fP}
\simeq \sum_{P}
\left(
\epsilon_{\alpha\beta}^{eP}
+ 3 \epsilon_{\alpha\beta}^{uP}
+ 3 \epsilon_{\alpha\beta}^{dP}
\right)$,
$n_f$ is the number density of $f$ in matter,
and we have taken into account the fact that the number density of
$u$ quarks and $d$ quarks are three times as that of
electrons.
The constraint on $\epsilon_{\alpha\beta}$ can be summarized as\,\cite{Davidson:2003ha,Biggio:2009nt}
\begin{eqnarray}
\left(
\begin{array}{ccc}
|\epsilon_{ee}|< 4\times 10^0 & |\epsilon_{e\mu}| < 3\times 10^{-1}
& |\epsilon_{e\tau}| < 3\times 10^0\\
& |\epsilon_{\mu\mu}| < 7\times 10^{-2}
& |\epsilon_{\mu\tau}| < 3\times 10^{-1}\\
& & |\epsilon_{\tau\tau}| < 2 \times 10^1
\end{array}
\right)~~\mbox{\rm at}~90\%\mbox{\rm CL}\,.
\label{b-eps-0}
\end{eqnarray}
From Eq. (\ref{b-eps-0}) we see that
%\begin{eqnarray}
$\epsilon_{e\mu} \simeq \epsilon_{\mu\mu} \simeq \epsilon_{\mu\tau} \simeq 0$
%\label{eps-mu}
%\end{eqnarray}
is satisfied.

\section{Parametrizations in solar and atmospheric neutrino analyses}
In Ref.\,\refcite{Fukasawa:2015jaa} it was shown that the atmospheric
neutrino measurements at Hyper-Kamiokande (HK) has a very good sensitivity
to the NSI, on the assumptions that (i) all the $\epsilon_{\alpha\mu}$ components of the NSI
vanish and (ii) the condition
\begin{eqnarray}
&{\ }&\hspace{-16mm}
\epsilon_{\tau\tau} = \frac{|\epsilon_{e\tau}|^2}{1 + \epsilon_{ee}}
\label{eq:mapping_fl0}
\end{eqnarray}
is satisfied, as is suggested by the high energy atmospheric
neutrino data.\,\cite{Friedland:2005vy}  In this talk we discuss the sensitivity of the
atmospheric neutrino measurements at HK to NSI without
the assumptions (i) and (ii) mentioned above.
In Ref.\,\refcite{Gonzalez-Garcia:2013usa},
the effect of NSI on solar neutrinos,
the $3 \times 3$ Hamiltonian in the Dirac equation
Eq.\,(\ref{sch1}) with the matter potential (\ref{matter-nsi}) is
reduced to an effective $2 \times 2$ Hamiltonian given by
\begin{eqnarray}
&{\ }&\hspace{-16mm}
H^{\rm eff}=
\frac{\Delta m^2_{21}}{4E}\left(\begin{array}{cc}
-\cos2\theta_{12} & \sin2\theta_{12}  \\
\sin2\theta_{12} & \cos2\theta_{12}
\end{array}\right) 
%\nonumber\\
%&{\ }&\hspace{4mm}
+
\left(\begin{array}{cc}
c^2_{13} A & 0 \\
0 & 0
\end{array}\right)  + 
 A\sum_{f=e,u,d} \frac{N_f}{N_e}
\left(\begin{array}{cc}
- \epsilon_{D}^f &  \epsilon_{N}^f \\
 \epsilon_{N}^{f*} &  \epsilon_{D}^f
\end{array}\right),
\end{eqnarray}
where  $\epsilon^f_{D}$ and $\epsilon^f_{N}$ are linear combinations of the standard NSI parameters:
\begin{eqnarray}
&{\ }&\hspace{-6mm}
\epsilon_{D}^f 
=
-\frac{c_{13}^2}{2}\left(\epsilon_{e e}^f-\epsilon_{\mu \mu}^f\right)+\frac{s_{23}^2-s_{13}^2c_{23}^2}{2}\left(\epsilon_{\tau \tau}^f-\epsilon_{\mu \mu}^f\right) 
\nonumber\\
&{\ }&\hspace{4mm}
+c_{13}s_{13}{\rm Re}\left[ e^{i\delta_{\rm CP}}\left(s_{23}\epsilon_{e \mu}^f
+c_{23}\epsilon_{e \tau}^f\right) \right]
%\nonumber\\
%&{\ }&\hspace{2mm}
-\left(1+s_{13}^2\right)c_{23}s_{23}{\rm Re}\left[\epsilon_{\mu \tau}^f\right]
\label{epsilond}\\
&{\ }&\hspace{-6mm}
\epsilon_{N}^f= -c_{13}s_{23}\epsilon_{e\tau}^f
\nonumber\\
&{\ }&\hspace{2mm}
+c_{13}c_{23}\epsilon_{e \mu}^f
+s_{13}c_{23}s_{23}e^{-i\delta_{\rm CP}}
\left(\epsilon_{\tau \tau}^f-\epsilon_{\mu \mu}^f\right) 
%\nonumber\\
%&{\ }&\hspace{2mm}
+s_{13}e^{-i\delta_{\rm CP}}\left( s_{23}^2\epsilon_{\mu\tau}^f-c_{23}^2\epsilon_{\mu\tau}^{f*} \right)
\,,
\label{epsilonn}
\end{eqnarray}
and $c_{jk}\equiv\cos\theta_{jk}$, $s_{jk}\equiv\sin\theta_{jk}$.
In the analysis of Ref.\,\refcite{Gonzalez-Garcia:2013usa},
one particular choice of $f=u$ or $f=d$ was taken at a time
because of the nontrivial composition profile of the Sun.

Since the parametrization which is
used in Ref.\,\refcite{Gonzalez-Garcia:2013usa} is different from the
one in (\ref{matter-std}) in the three flavor basis, a non-trivial mapping
is required to compare the results in these two parametrizations.
It is instructive to see how the two parametrizations
are related.  Here let us assume for simplicity that
$\theta_{13}=0$, $\theta_{23}=\pi/4$, $\epsilon_{\alpha\mu}=0$
and $\epsilon_{\tau\tau} = |\epsilon_{e\tau}|^2/(1 + \epsilon_{ee})$,
although in our numerical analysis we do not assume these
conditions.
Then, noting that $\epsilon_{\alpha\beta}=3\epsilon_{\alpha\beta}^d$ and
$\epsilon_{D}=\epsilon_{D}^d$,
$\epsilon_{N}=\epsilon_{N}^d$,
Eqs.\,(\ref{epsilond}) and (\ref{epsilonn}) become
\begin{eqnarray}
&{\ }&
\hspace*{-55mm}
3\epsilon_{D} = -\frac{1}{2}\epsilon_{e e}
+\frac{|\epsilon_{e\tau}|^2}{4(1 + \epsilon_{ee})}\,,\quad
%\label{eq:mapping_fl1}\\
%&{\ }&
%\hspace*{-60mm}
3\epsilon_{N} = -\frac{1}{\sqrt{2}}\,\epsilon_{e\tau}\,.
\label{eq:mapping_fl1}
\end{eqnarray}
From Eq.\,(\ref{eq:mapping_fl1}) we have
\begin{eqnarray}
&{\ }&
\hspace*{-25mm}
\epsilon_{e\tau} = -3\sqrt{2}\epsilon_{N}\,,\quad
\epsilon_{e e} = -\frac{1}{2} -3 \epsilon_{D}
+\left\{(\frac{1}{2} -3 \epsilon_{D})^2+|3\epsilon_{N}|^2\right\}^{1/2}\,,
\label{eq:mapping_fl2}
\end{eqnarray}
where the sign in the solution in the quadratic equation for
$\epsilon_{e e}=0$
was chosen so that the standard case ($\epsilon_{D}=\epsilon_{N}=0$)
is reduced to $\epsilon_{e e}=0$.

\begin{figure}
\hspace{-0.2in}
\includegraphics[scale=0.7]{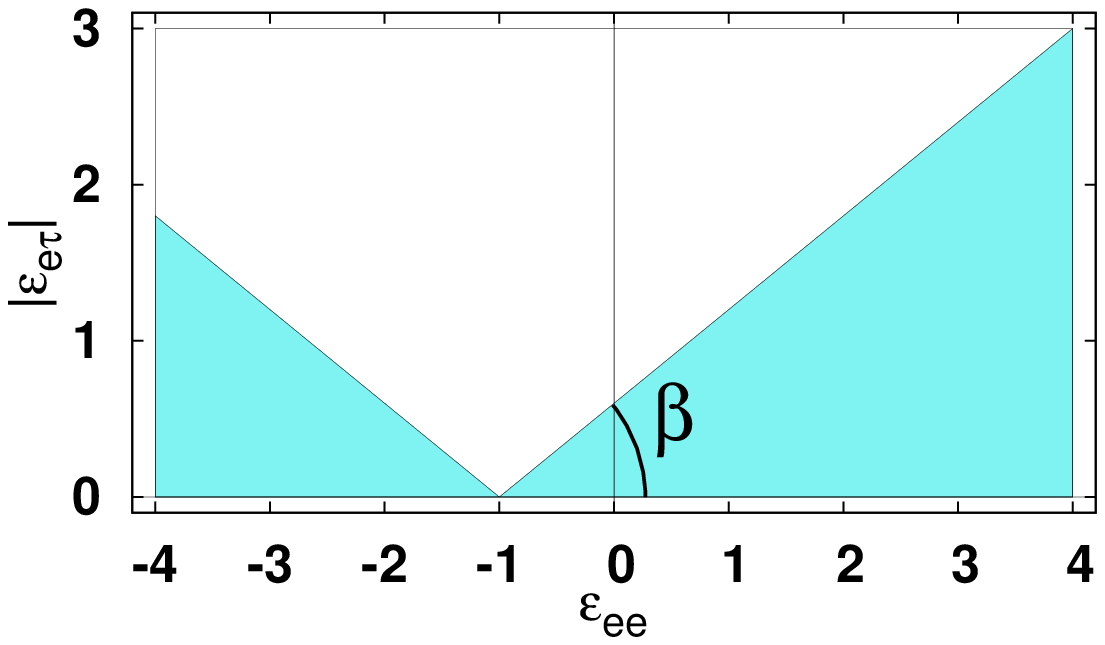}
\hspace{-0.5in}
\includegraphics[scale=0.7]{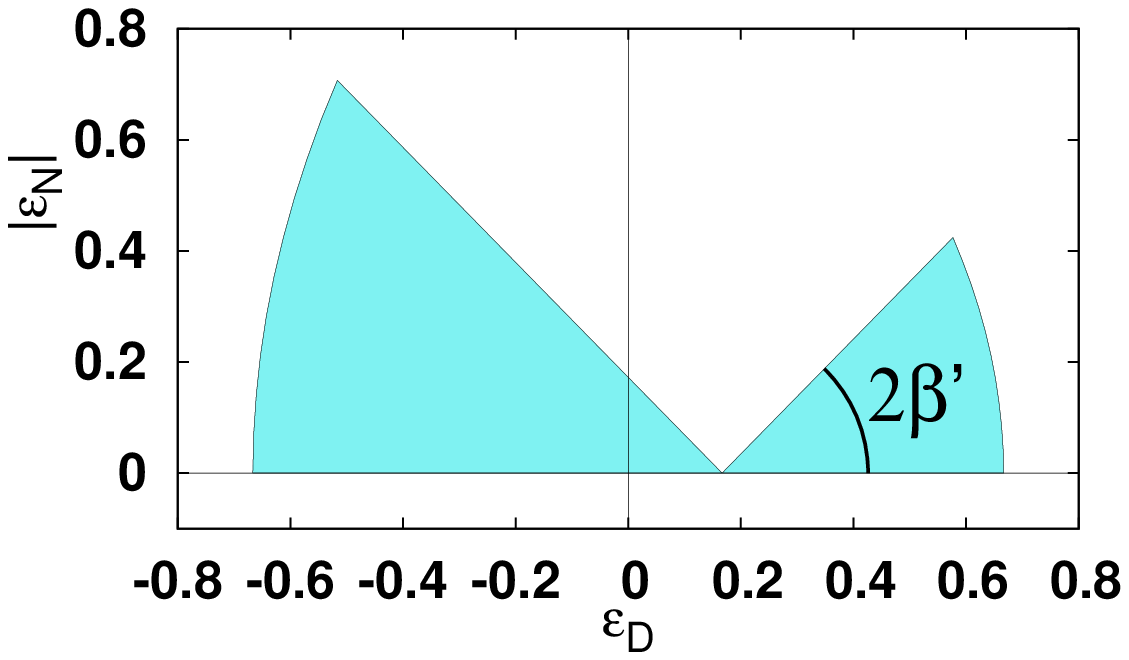}
\caption{
The current allowed regions in the ($\epsilon_{ee}$, $|~\epsilon_{e\tau}|$)
plane (left panel) and in the ($\epsilon_D$, $\epsilon_N$) (right panel).
The region $\epsilon_D > 1/6$ ($< 1/6$) in the right panel corresponds to
$\epsilon_{ee} < -1$ ($ > -1$) in the left panel.
The left (right) edge $\epsilon_{ee} = -4$ ($\epsilon_{ee} = 4$) 
in the left panel corresponds to the quadratic curve on the
right (left) end in the right panel.
}
\label{fig:fig0}
\end{figure}

The matter potential (\ref{matter-nsi})
with $\epsilon_{\alpha\mu}=0$ and
$\epsilon_{\tau\tau} = |\epsilon_{e\tau}|^2/(1 + \epsilon_{ee})$
can be diagonalized as
\begin{eqnarray}
&{\ }&
\hspace*{-30mm}
A\left(
\begin{array}{ccc}
1+ \epsilon_{ee}~~ & 0 & \epsilon_{e\tau}\\
0 & 0 & 0\\
\epsilon_{e\tau}^\ast & 0 & ~~|\epsilon_{e\tau}|^2/(1 + \epsilon_{ee})
\end{array}
\right)
\nonumber\\
&{\ }&
\hspace*{-35mm}
=A\,e^{i\gamma\lambda_9}e^{-i\beta\lambda_5}
\mbox{\rm diag}
\left\{
%1+\epsilon_{ee}+\frac{|\epsilon_{e\tau}|^2}{1+\epsilon_{ee}},0,0
1+\epsilon_{ee}+|\epsilon_{e\tau}|^2/(1+\epsilon_{ee}),0,0
\right\}\,
e^{i\beta\lambda_5}e^{-i\gamma\lambda_9},
\label{matter-nsi2}
\end{eqnarray}
where
% $\gamma\equiv(1/2)$arg($\epsilon_{e\tau}$) and
\begin{eqnarray}
&{\ }&
\hspace*{-5mm}
\tan\beta\equiv
\frac{\left|\epsilon_{e\tau}\right|}
{1+\epsilon_{ee}}\,,~~
\gamma\equiv\frac{1}{2}\mbox{\rm arg}\,(\epsilon_{e\tau})\,,~~
\lambda_5\equiv\left(
\begin{array}{ccc}
0&0&-i\cr
0&0&0\cr
i&0&0
\end{array}\right)\,,~~
\lambda_9\equiv\left(
\begin{array}{ccc}
1&0&0\cr
0&0&0\cr
0&0&-1
\end{array}\right)\,.
\nonumber
\end{eqnarray}
The quantity $\tan\beta$ which is expressed by
the matter angle $\beta$ can be regarded as
the gradient of the straight line 
$|\epsilon_{e\tau}|=(1+\epsilon_{ee})\tan\beta$
as we can see in the left panel in Fig.\ref{fig:fig0}.

In the present case, from (\ref{eq:mapping_fl2}) we have
\begin{eqnarray}
&{\ }&
\hspace*{-55mm}
\tan\beta
=\frac{\left|3\sqrt{2}\epsilon_N\right|}
{
1/2-3\epsilon_{D}
+\left\{\left(1/2-3\epsilon_{D}
\right)^2
+|3\epsilon_N|^2
\right\}^{1/2}
}\,.
\label{eq:mapping_fl5}
\end{eqnarray}
Now if we introduce a new angle
\begin{eqnarray}
&{\ }&
\hspace*{-33mm}
\tan\beta'\equiv\frac{\tan\beta}{\sqrt{2}}
=
\frac{\left|3\epsilon_N\right|}
{
1/2-3\epsilon_{D}
+\left\{\left(1/2-3\epsilon_{D}
\right)^2
+|3\epsilon_N|^2
\right\}^{1/2}
}\,,
\label{eq:mapping_fl6}
\end{eqnarray}
then from Eq.\,(\ref{eq:mapping_fl6}) we obtain
\begin{eqnarray}
&{\ }&
\hspace*{-70mm}
\tan2\beta'=
\frac{2\tan\beta'}
{1-\tan^2\beta'}
=\frac{\left|3\epsilon_N\right|}
{1/2-3\epsilon_{D}}\,.
\label{eq:mapping_fl7}
\end{eqnarray}
Eq.\,(\ref{eq:mapping_fl7}) implies that
the allowed region of the atmospheric
neutrino experiment with
the parabolic relation
(\ref{eq:mapping_fl0}) is approximately
the one
surrounded by the $\epsilon_N=0$ axis
and the straight line 
$|\epsilon_N|=|\tan2\beta'||1/6-\epsilon_{D}|$
with the gradient $|\tan2\beta'|$ and 
the $x$-intercept $\epsilon_D=1/6$.

The constraint from the atmospheric neutrino experiments
can be expressed as\,\cite{Fukasawa:2015jaa}
\begin{eqnarray}
&{\ }&\hspace{-80mm}
\left|\frac{\epsilon_{e\tau}}{1+\epsilon_{ee}}\right|
\lesssim 0.6\quad\mbox{\rm at}~90\%\mbox{\rm CL}\,.
\label{tanb}
\end{eqnarray}
Combining the constraints (\ref{b-eps-0}) and
(\ref{tanb}), the
current allowed region, without assuming that the NSI
accounts for the solar neutrino and KamLAND data,
is approximately given by the
shaded area in the ($\epsilon_{ee}$, $|\epsilon_{e\tau}|$)-plane
(the ($\epsilon_{D}$, $|\epsilon_{N}|$)-plane)
in left (right) panel in Fig.\ref{fig:fig0}.

\section{Results}
\begin{figure}
\includegraphics[scale=0.42]{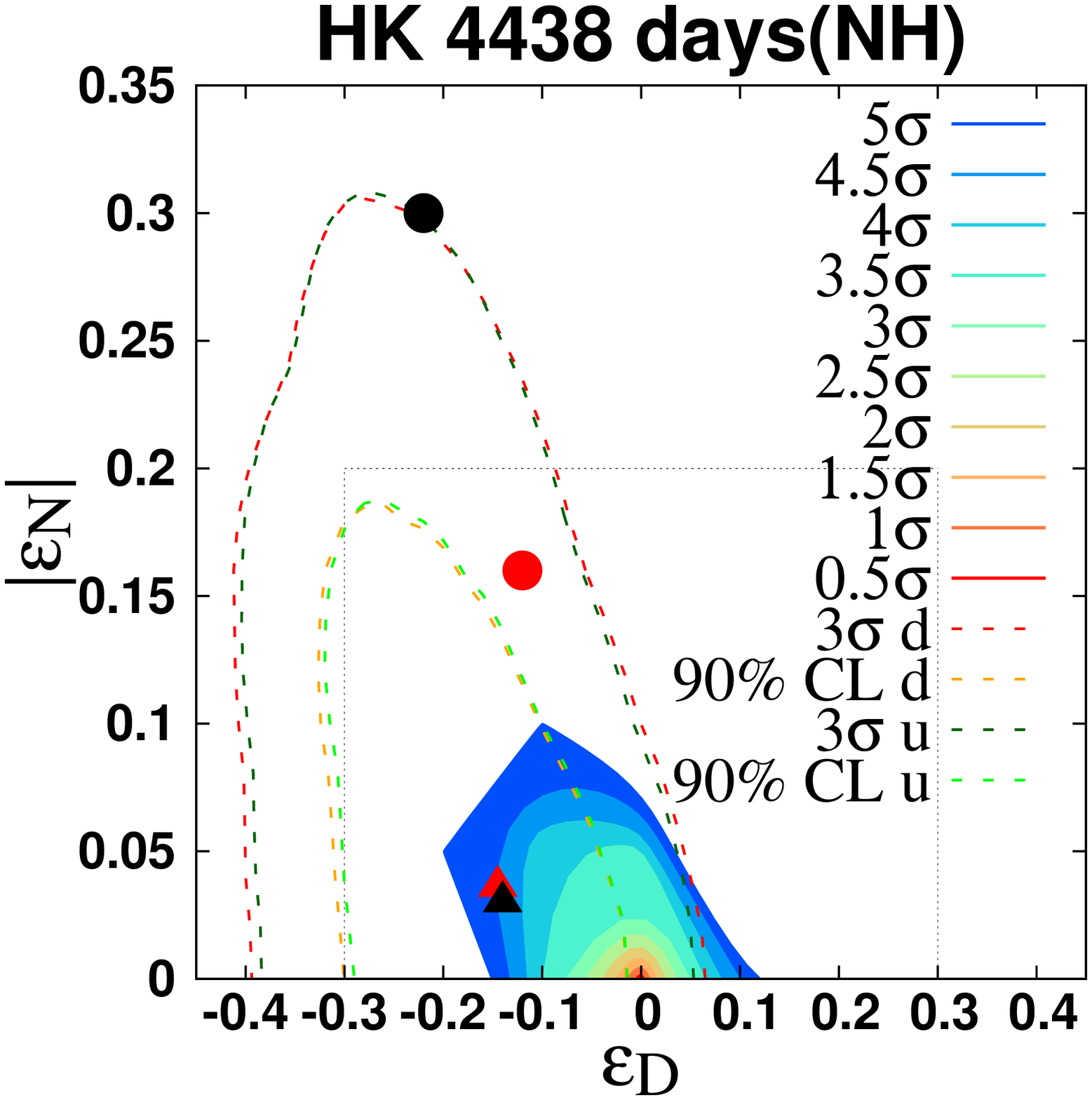}
\includegraphics[scale=0.42]{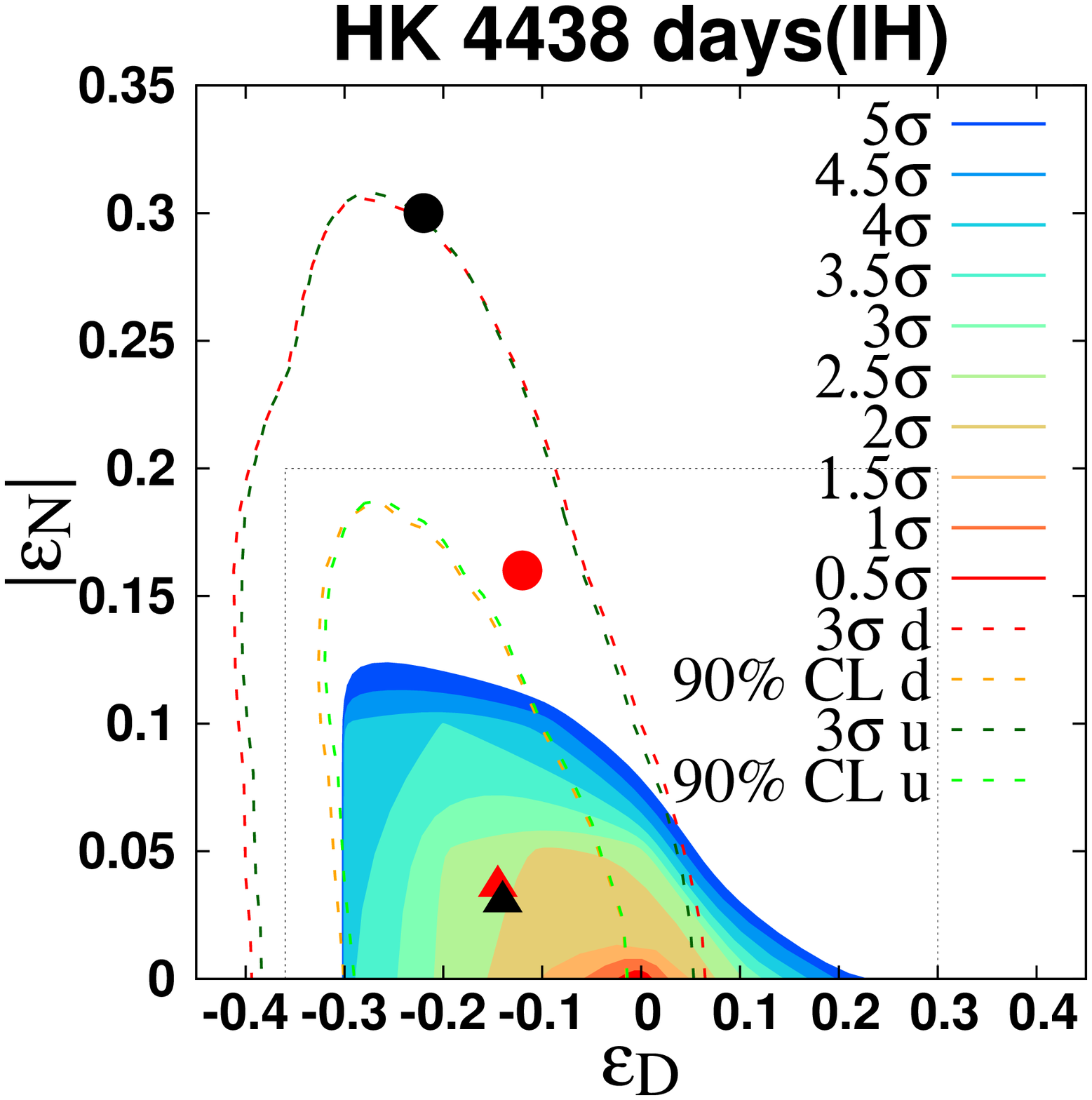}
\caption{
The allowed region in the ($\epsilon_{D}$, $|\epsilon_{N}|$) plane from
the HK atmospheric neutrino data for the normal hierarchy (left panel) and for the inverted hierarchy (right panel).
We calculated $\chi^2$ for ($\epsilon_{D}$, $|\epsilon_{N}|$) inside
the area surrounded by dotted lines and at the best fit points.
The red ($f=d$) and black ($f=u$) circles indicate the best fit point
from the solar neutrino and KamLAND data for NSI with
$(\epsilon_{D}^d,\epsilon_{N}^d)=(-0.12,-0.16)$ (red) and that for NSI
with $(\epsilon_{D}^u,\epsilon_{N}^u)=(-0.22,-0.30)$ (black),
respectively.
The red and black triangles indicate the best fit value from the global neutrino oscillation experiments analysis for NSI with $(\epsilon_{D}^d,\epsilon_{N}^d)=(-0.145,-0.036)$ (red) and that for NSI with $(\epsilon_{D}^u,\epsilon_{N}^u)=(-0.140,-0.030)$ (black), respectively.
The dashed lines are the boundaries of the allowed regions from the global neutrino oscillation experiments analysis.
For reference, we plotted for both the cases with $f=u$ and $f=d$.
}
\label{fig:fig1}
\end{figure}

We performed a $\chi^2$ analysis of the HK atmospheric
neutrino experiment, assuming that
HK measures the atmospheric neutrinos with
the fiducial volume 0.56 Mton for 20 years.
We also assumed that
the experimental numbers of events are those
with the standard three flavor oscillation
parameters.  From the deviation from
the numbers of events with the standard oscillation
scenario we have obtained the allowed
region for NSI in the ($\epsilon_{D}$, $|\epsilon_{N}|$) plane.
The results are shown in Fig.\,\ref{fig:fig1}
in the case of both mass hierarchies, assuming that we know
the mass hierarchy.\,\footnote{
The details of our analysis can be found
in Ref.\,\refcite{Fukasawa:2015jaa}.}
The best fit values $(\epsilon_{D}^d,\epsilon_{N}^d)=(-0.12,-0.16)$ for NSI with $f=d$ from the solar neutrino and KamLAND data given by Ref.\,\refcite{Gonzalez-Garcia:2013usa} is excluded at $11\sigma$ ($8.2\sigma$) for the normal (inverted) hierarchy.
In the case of NSI with $f=u$, the best fit value $(\epsilon_{D}^u,\epsilon_{N}^u)=(-0.22,-0.30)$ is far from the standard scenario $(\epsilon_{D},\epsilon_{N})=(0.0,0.0)$ compared with the case of $f=u$ and also excluded at $38\sigma$ ($35\sigma$) for the normal (inverted) hierarchy.
On the other hand, the best fit value from the global analysis of the neutrino oscillation data \cite{Gonzalez-Garcia:2013usa} 
$(\epsilon_{D}^d,\epsilon_{N}^d)=(-0.145,-0.036)$ for NSI with $f=d$
is excluded at $5.0\sigma$ ($3.7\sigma$)  for the normal (inverted) hierarchy.
In the case of NSI with $f=u$, the best fit value 
$(\epsilon_{D}^u,\epsilon_{N}^u)=(-0.140,-0.030)$
is excluded at $5.0\sigma$ ($1.4\sigma$) 
for the normal (inverted) hierarchy.

\section{Conclusion\label{conclusion}}
In this talk we have presented the sensitivity
of the future HK atmospheric neutrino
experiment to NSI which is suggested by
the tension between the mass squared differences
from the solar neutrino and KamLAND data.
If there are no non-standard interactions in
nature, then the best fit point
of the combined analysis 
of the solar neutrino and KamLAND data by 
Ref.\,\refcite{Gonzalez-Garcia:2013usa} can be excluded at
more than $11\sigma$ ($8\sigma$) in the case of the normal
(inverted) hierarchy, while the best fit point
of the global analysis in Ref.\,\refcite{Gonzalez-Garcia:2013usa}
can be excluded at
$5.0\sigma$  ($1.4\sigma$) in the case of the normal
(inverted) hierarchy.

In the HK experiment, because of
large statistics, it is expected that
the solar neutrino observation,
whose typical energy is low ($E_\nu~\sim$ several MeV),
can test the tension between
the solar and KamLAND data by the day night
effect.\,\cite{kajita:2016now}
On the other hand, our result
indicates that
this tension can be tested also
by the atmospheric
neutrino observation,
whose typical energy is high
($E_\nu\sim{\cal O}$(10) GeV),
through the matter effect at the same HK
facility.

\section*{Acknowledgments}
This research was partly supported by a Grant-in-Aid for Scientific
Research of the Ministry of Education, Science and Culture, under
Grants No. 25105009, No. 15K05058, No. 25105001 and No. 15K21734.

\end{document}